\documentclass[usenatbib,usegraphicx]{mn2e}
\def\msun{{\rm M}_{\odot}}
\def\rsun{{\rm R}_{\odot}}

\begin{document}

\title[Rotation periods of low-mass stars in M50]{The Monitor project:
    Rotation periods of low-mass stars in M50}
\author[J.~Irwin et al.]{Jonathan~Irwin$^{1,2}$\thanks{E-mail: jirwin at
    cfa.harvard.edu},
    Suzanne~Aigrain$^{3}$, Jerome~Bouvier$^{4}$, Leslie~Hebb$^{5}$,
\newauthor
Simon~Hodgkin$^{1}$, Mike~Irwin$^{1}$, Estelle~Moraux$^{4}$ \\
$^{1}$Institute of Astronomy, University of Cambridge, Madingley Road,
  Cambridge, CB3 0HA, United Kingdom \\
$^{2}$Harvard-Smithsonian Center for Astrophysics, 60 Garden Street,
    Cambridge, MA 02138, USA \\
$^{3}$Astrophysics Group, School of Physics, University of Exeter, Stocker Road,
  Exeter, EX4 4QL, United Kingdom \\
$^{4}$Laboratoire d'Astrophysique, Observatoire de Grenoble, BP 53,
  F-38041 Grenoble C\'{e}dex 9, France \\
$^{5}$School of Physics and Astronomy, University of St Andrews,
  North Haugh, St Andrews, KY16 9SS, Scotland}
\date{}

\maketitle

\begin{abstract}
We report on the results of a time-series photometric survey of M50
(NGC 2323), a $\sim 130\ {\rm Myr}$ open cluster, carried out using
the CTIO 4m Blanco telescope and Mosaic-II detector as part of the
Monitor project.  Rotation periods were derived for $812$ candidate
cluster members over the mass range $0.2 \la M/\msun \la 1.1$.  The
rotation period distributions show a clear mass-dependent morphology,
statistically indistinguishable from those in NGC 2516 and M35 taken
from the literature.  Due to the availability of data from three
observing runs separated by $\sim 10$ and $1\ {\rm month}$ timescales,
we are able to demonstrate clear evidence for evolution of the
photometric amplitudes, and hence spot patterns, over the $10\ {\rm
month}$ gap, although we are not able to constrain the timescales for
these effects in detail due to limitations imposed by the large gaps
in our sampling, preventing use of the phase information.
\end{abstract}

\section{Introduction}
\label{intro_section}

M50 is a populous ($\sim 2100$ stars brighter than $V \sim 23$;
\citealt{kalirai03}) open cluster of comparable age to the Pleiades,
at moderate distance.  We adopt the parameters of \citet{kalirai03}
for the remainder of this work: a main sequence turn-off age of $130\
{\rm Myr}$, distance $1000^{+81}_{-75}\ {\rm pc}$ and reddening
$E(B-V) = 0.22\ {\rm mag}$ (an average of values from
\citealt*{claria98} and \citealt{hoag67}; see \citealt{kalirai03}).
For comparison, the classical main sequence turn-off age for the
Pleiades is $100\ {\rm Myr}$ \citep*{mmm93}.

Despite having a very favourable angular size for CCD observations,
there are only two previous CCD-based studies of the cluster
population in the literature, from \citet{kalirai03} and
\citet{sharma06}, obtaining similar values for the cluster parameters.
We have undertaken a time-domain photometric survey of M50 with the
joint aims of searching for eclipses due to stellar, brown dwarf and
planetary companions orbiting low-mass cluster member stars, and to
obtain a large sample of rotation periods to study the evolution of
stellar angular momentum in conjunction with previous measurements in
the literature.

In addition to studying angular momentum evolution, we can also begin
to constrain the effects of other cluster parameters such as metallicity
and environment on the rotation period distributions, by comparing the
M50 distribution with data for clusters of comparable age.  Large
samples of period measurements are available in NGC 2516 ($\sim 150\
{\rm Myr}$; \citealt*{jth2001}) from \citet{i2007b}, covering $0.15
\la M/\msun \la 0.7$, and M35 (also $\sim 150\ {\rm Myr}$;
\citealt*{barrado01}) from \citet{meibom08}, covering masses down to
$\sim 0.6\ \msun$.  Although there are relatively few rotation periods
measured, especially for M-dwarfs, in the Pleiades, a large sample of
$v \sin i$ measurements are available covering masses down to
$\sim 0.3\ \msun$ (\citealt{plev1}; \citealt{plev2}; \citealt{plev3};
\citealt{plev4}; \citealt{plev5}; \citealt{plev6}), which can be
compared to the rotation periods in a statistical sense.

The rotational evolution of low-mass stars on the pre--main-sequence
(PMS) is dominated by stellar contraction, with angular momentum
thought to be regulated by processes relating to the star-disc
interaction (e.g. accretion-driven winds; \citealt{mp05}, or ``disc
locking''; \citealt{k91}; \citealt*{cc95}), giving rise to a spread in
rotation rates determined predominantly by the disc lifetime
(e.g. \citealt{k97}; \citealt*{bfa97}; \citealt*{spt00}).  As the
stars arrive on the zero age main sequence (ZAMS), the stellar
contraction ceases, and angular momentum losses via magnetised stellar
winds dominate the subsequent evolution.  Observations indicate that
the rotation rates of solar-type stars between the age of the Hyades
($\sim 625\ {\rm Myr}$; \citealt{perry98}) and the age of the Sun are
well-described by $\omega \propto t^{-1/2}$, the famous \citet{sk72}
law (e.g. \citealt{soderblom83}).  This can be reproduced in a more
theoretically-motivated framework from parametrised angular momentum
loss laws (usually based on \citealt{k88}).

Observations of the Pleiades and other young clusters at $\sim 50-100\
{\rm Myr}$ represent a ``snapshot'' of the rotational evolution
process at the point where solar-type stars have recently reached the
ZAMS, and before significant angular momentum losses due to stellar
winds have taken place.  These indicate that although the evolution of
the slowest rotators from this age to the age of the Hyades (and the
Sun) can be reproduced by the \citet{sk72} law, these clusters show a
spread in rotation rates at a given mass, and a number of ``ultrafast
rotators'' (e.g. \citealt{plev3}), neither of which are seen in the
Hyades and at older ages: here the rotation rate is typically found to
follow a fairly well-defined function of mass
(e.g. \citealt{radick87}).  In order to reproduce the ultrafast
rotators on the ZAMS, most modellers modify the \citet{k88} formalism
to incorporate saturation of the angular momentum losses above a
critical angular velocity $\omega_{\rm sat}$ (\citealt{plev2};
\citealt{bs96}).  The saturation is further assumed to be
mass-dependent, to account for the mass-dependent spin-down timescales
observed on the early main sequence.

There is mounting evidence in the literature that this picture of
angular momentum evolution is still not sufficient to reproduce the
observations on the ZAMS, in particular over the interval between
$\sim 50-500\ {\rm Myr}$, if we assume the stars rotate as solid
bodies.  The observations indicate that such models produce a
spin-down over this age range that is too rapid, particularly for the
slowest rotators in open clusters (e.g. \citealt{i2007b}).  Several
studies (e.g. \citealt{k97}; \citealt{a98}; \citealt{i2007b}) have
invoked core--envelope decoupling, where the radiative core and
convective envelope of the star are allowed to have different rotation
rates, as a means to produce a more shallow evolution, by coupling
angular momentum from a rapidly-spinning core (which experiences
little angular momentum loss assuming the disc and wind couple
predominantly to the outer convective regions of the star) to the
convective envelope on timescales of a few $100\ {\rm Myr}$ to provide
a ``late time replenishment'' of the surface rotation rate.

Open clusters of $50-500\ {\rm Myr}$ age represent an ideal
testing ground for these models.  The rotational evolution is 
strongly mass-dependent, so in order to decouple the mass effect
from the time-dependence, extremely large sample sizes are required,
such that we can obtain good statistics over small bins in mass.  M50
represents an ideal target for such a study, having a large population
over a relatively small area of sky, which can be observed very
efficiently using the multiplex advantage afforded by a standard
wide-field CCD mosaic.

The remainder of the paper is structured as follows: the observations
and data reduction are described in \S \ref{odr_section}, and the
colour magnitude diagram (CMD) of the cluster and candidate membership
selection are presented in \S \ref{memb_section}.  The method we use
for obtaining photometric periods is summarised in \S
\ref{period_section} (see \citealt{i2006} for a more detailed
discussion).  Our results are given in \S \ref{results_section}, and
\S \ref{conclusions_section} summarises our conclusions.

\section{Observations and data reduction}
\label{odr_section}

Photometric monitoring observations were obtained as part of the
Monitor project \citep{a2007}, using the 4m 
Blanco telescope at Cerro Tololo Interamerican Observatory (CTIO),
with the Mosaic-II imager.  This instrument provides
a field of view of $\sim 36' \times 36'$ ($0.37\ {\rm sq. deg}$),
using a mosaic of eight $2048 \times 4096$ pixel CCDs, at a
scale of $\sim 0.27'' / {\rm pix}$.  A total of $\sim 95\ {\rm hours}$
of photometric monitoring was conducted, spread over three distinct
observing runs: $6 \times$ full--nights in two three-night segments
spanning 2005 Feb 04--06 and 2005 Feb 14--16, $8 \times 1/2$--nights
between 2005 Dec 24 and 2006 Jan 06, and $4 \times$ full-nights
between 2006 Jan 28 and 2006 Feb 01.  M50 was observed in a single
telescope pointing centred on the cluster, for $\sim 8\ {\rm hours}$
per night ($\sim 4\ {\rm hours}$ per $1/2$-night), in parallel with
another field in the cluster NGC 2362, the results from which were
published in \citet{i2008b}.  Exposure times were $75\ {\rm s}$ in
$i$-band, giving a cadence of $\sim 6\ {\rm minutes}$ (composed of $2
\times$ $75\ {\rm s}$ exposures plus $2 \times$ $100\ {\rm s}$ readout
time, slewing between M50 and NGC 2362 during readout).  We also
obtained deep $V$-band exposures ($2 \times 600\ {\rm s}$, $450\ {\rm
  s}$ and $150\ {\rm s}$) which were stacked and used to produce a
colour-magnitude diagram of the cluster.

\begin{figure}
\centering
\includegraphics[angle=270,width=3.2in]{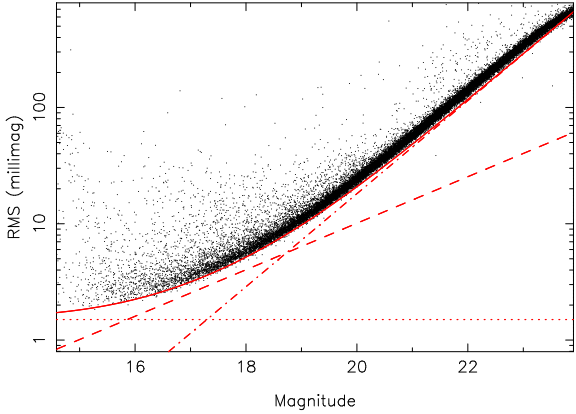}

\caption{Plot of RMS scatter per data point (measured over the entire
  data-set) as a function of magnitude
  for the $i$-band observations of a single field in M50, for all
  unblended objects with stellar morphological classifications.  The
  diagonal dashed line shows the expected RMS from Poisson noise in
  the object, the diagonal dot-dashed line shows the RMS from sky
  noise in the photometric aperture, and the dotted line shows an
  additional $1.5\ {\rm mmag}$ contribution added in quadrature to
  account for systematic effects.  The solid line shows the overall
  predicted RMS, combining these contributions.}

\label{rmsplot}
\end{figure}

For a full description of our data reduction steps, the reader is
referred to \citet{i2007a}.  Briefly, we used the pipeline for
the INT wide-field survey \citep{il2001} for 2-D 
instrumental signature removal (bias correction, flatfielding,
defringing) and astrometric and photometric calibration.  We then 
generated a master catalogue for each filter by stacking $20$
of the frames taken in the best conditions (seeing, sky brightness and
transparency) and running the source detection software on the stacked
image.  The resulting source positions were used to perform aperture
photometry on all of the time-series images.  We achieved a per data
point photometric precision of $\sim 2-4\ {\rm mmag}$ for the
brightest objects, with RMS scatter $< 1$ per cent for $i \la 19$ (see
Figure \ref{rmsplot}).  A signal-to-noise ratio of $5$ (corresponding
approximately to the detection limit for point sources on a single
frame of the differential photometry) is reached at $i \sim 22.7$.

Our source detection software flags any objects detected as having
overlapping isophotes.  This information is used, in conjunction with
a morphological image classification flag also generated by the
pipeline software \citep{il2001} to allow us to identify non-stellar
or blended objects in the time-series photometry.

Photometric calibration of our data was carried out using regular
observations of \citet{l92} equatorial standard star fields in the
usual way.

Light curves were extracted from the data for $\sim 63\,000$ objects,
$42\,000$ of which had stellar morphological classifications ($\sim
23$ per cent of these are flagged as having overlapping isophotes by
the source detection software, and thus may be blended), using our standard
aperture photometry techniques, described in \citet{i2007a}. We fit a
2-D quadratic polynomial to the residuals in each frame  (measured for
each object as the difference between its magnitude on the frame in
question and the median calculated across all frames) as a function of
position, for each of the $8$ CCDs separately. Subsequent removal of
this function accounts for effects such as varying differential
atmospheric extinction across each frame.  Over a single CCD, the
spatially-varying part of the correction remains small, typically
$\sim 0.02\ {\rm mag}$ peak-to-peak.  The reasons for using this
technique are discussed in more detail in \citet{i2007a}.

For the production of deep CMDs, we stacked 20 $i$-band observations,
taken in good seeing and photometric conditions, and all of the
$V$-band observations.  The limiting magnitudes on these stacked
images, measured as the approximate magnitude at which our catalogues
are $50$ per cent complete, were $V \simeq 24.4$ and $i \simeq 23.6$.

\section{Selection of candidate low-mass members}
\label{memb_section}

\subsection{The $V$ versus $V - I$ CMD}
\label{cmd_section}

\begin{figure}
\centering
\includegraphics[angle=270,width=3.5in]{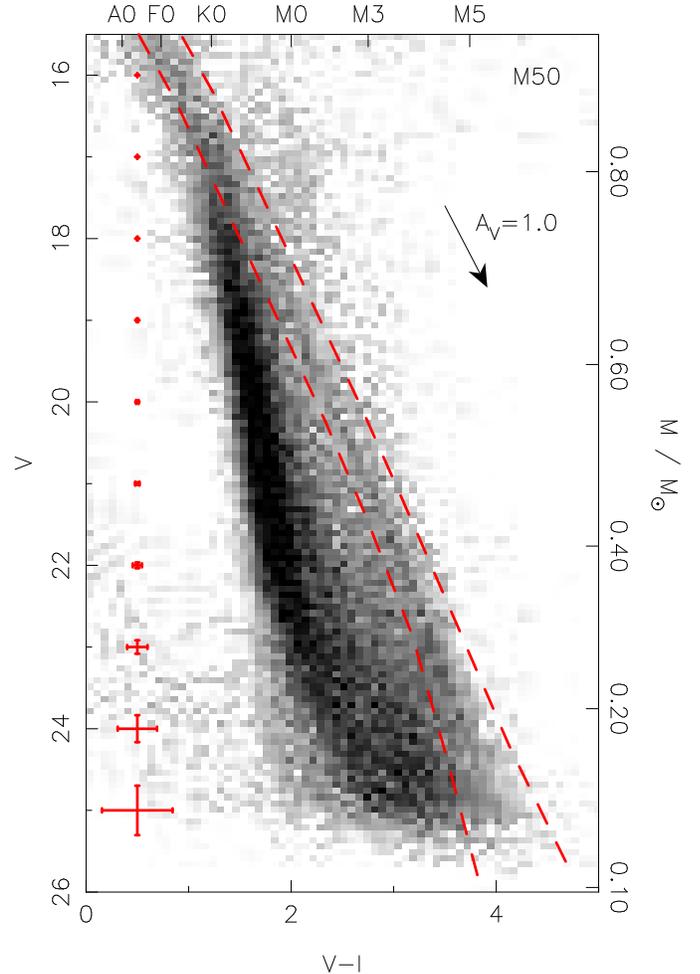}

\caption{$V$ versus $V - I$ CMD of M50 from stacked images plotted as
  a ``Hess diagram'' (greyscale map of the number density of sources
  in $0.1\ {\rm mag}$ bins), for
all objects with stellar morphological classification.  The cluster
sequence is clearly visible on the right-hand side of the diagram.
The boundaries of the region used to select photometric candidate
members are shown by the dashed lines (all objects between the dashed
lines were selected).  The reddening vector for $A_V = 1.0$ is shown
at the right-hand side of the diagram.  The mass scale is from the
NextGen models of \citet{bcah98}, interpolated to $130\ {\rm Myr}$,
using our empirical isochrone to convert the $V$ magnitudes to $I$
magnitudes, and subsequently obtaining the masses from these, due to
known problems with the $V$ magnitudes from the models (see, for
example, \citealt{bcah98}).  The error bars at the left-hand side of
the plot indicate the typical photometric error for an object on the
cluster sequence.}

\label{cmd}
\end{figure}

Our CMD of M50 is shown in Figure \ref{cmd}.  The $V$ and $i$
measurements were converted to the standard Johnson-Cousins
photometric system using colour equations derived from our standard
star observations:
\begin{eqnarray}
(V - I)& = &(V_{\rm ccd} - i_{\rm ccd})\ /\ 0.899 \\
V& = &V_{\rm ccd} + 0.005\ (V - I) \\
I& = &i_{\rm ccd} - 0.096\ (V - I)
\end{eqnarray}

Candidate cluster members were selected by defining an empirical cluster
sequence `by eye' to follow the clearly-visible cluster single-star
sequence.  The cuts were defined by moving this line along a vector
perpendicular to the cluster sequence, by amounts $k - \sigma(V -
I)$ and $k + \sigma(V - I)$ as measured along this vector, where
$\sigma(V - I)$ is the photometric error in the $V - I$ colour.  The
values of $k$ used were $-0.125\ {\rm mag}$ for the lower line and
$0.25\ {\rm mag}$ for the upper line on the diagram, making the
brighter region wider to avoid rejecting binary and multiple systems,
which are overluminous for their colour compared to single stars.
$4249$ candidate photometric members were selected, over the full $V$
magnitude range from $V = 15.5$ to $26$, but the cluster sequence
becomes difficult to distinguish from the field population for $V \ga
24$ ($M \la 0.2\ \msun$).

In this and subsequent sections, we make use of mass estimates for the
cluster members.  These were derived from the $I$-band absolute
magnitudes and the models of \citet{bcah98}, interpolated to $130\
{\rm Myr}$ age.  We did not use the $V - I$ colour due to known
problems with the $V$-band magnitudes from the models (see
\citealt{bcah98}).

\subsection{Contamination}
\label{contam_section}

Since the cluster sequence is not well-separated from the field in the
CMD, it is important to estimate the level of field star contamination
in the sample of candidate cluster members.  In this work, we have
used Galactic models to obtain a simple first estimate of the
contamination level.  This will be refined using spectroscopic
follow-up observations in a later publication.

\begin{figure}
\centering
\includegraphics[angle=270,width=3in]{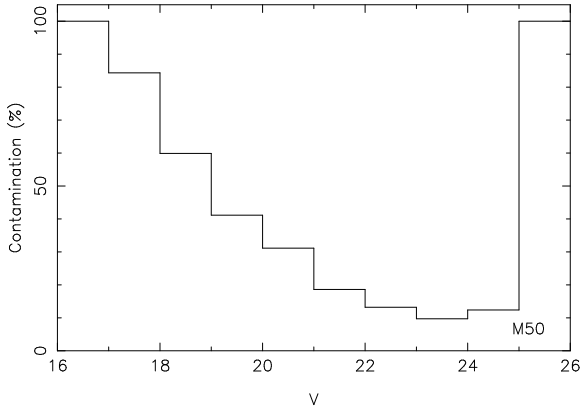}

\caption{Contamination estimated from Galactic models, measured as the
  ratio of the calculated number of objects in each magnitude bin from
  the models, to the number of objects detected and classified as
  candidate cluster members in that magnitude bin.  Note that bins
  with contamination estimates $> 100$ per cent (where there were more
  objects in that bin from the Galactic model than were actually
  observed) have been truncated to $100$ per cent.}

\label{contam}
\end{figure}

The Besan\c{c}on Galactic models \citep{r2003} were used to to
generate a simulated catalogue of objects passing our selection
criteria at the Galactic coordinates of M50 ($l = 221.7^\circ$,
$b = -1.3^\circ$), covering the total FoV of $\sim 0.35\ {\rm sq.deg}$
(including gaps between detectors).  The same selection process as
above for the cluster members was applied to this catalogue to find
the contaminant objects.  A total of $1970$ simulated objects passed
these membership selection criteria, giving an overall contamination
level of $\sim 47$ per cent after correcting for bins where the number
of objects predicted by the models exceeded the number actually
observed (we simply assumed $100$ per cent field contamination in
these bins). Figure \ref{contam} shows the contamination as a function
of $V$ magnitude.  Note that this figure should be treated with
extreme caution due to the need to use Galactic models, and especially
given the overestimation of the numbers of observed objects by the
models.

\section{Period detection}
\label{period_section}

\subsection{Method}
\label{method_section}

The method we use for detection of periodic variables was described in
detail in \citet{i2006}.  It uses least-squares fitting of
sine curves to the time series $m(t)$ (in magnitudes) for all
candidate cluster members, using the form:
\begin{equation}
m(t) = m_{\rm dc} + \alpha \sin(\omega t + \phi)
\label{sine_eqn}
\end{equation}
where $m_{\rm dc}$ (the DC light curve level), $\alpha$ (the
amplitude) and $\phi$ (the phase) are free parameters at each value of
$\omega$ over an equally-spaced grid of frequencies, corresponding to
periods from $0.05 - 50\ {\rm days}$ for the present data-set.

For M50, we modified this to fit separate coefficients $m_{\rm dc}$,
$\alpha$ and $\phi$ for each of the three observing runs (a total of
$9$ coefficients at each value of $\omega$) to allow for changes in
the spot patterns that give rise to the observed photometric
modulations.  The period was required to remain the same over all the
observing runs, which is expected since it should represent the
underlying rotation period of the star, and this will change by a
negligible amount due to rotational evolution over only $1\ {\rm
  yr}$.  This procedure is necessary for such a long data-set since
significant evolution of the spot coverage of our targets is expected
over timescales shorter than $1\ {\rm yr}$.

Periodic variable light curves were selected as before by evaluating
the change in reduced $\chi^2$:
\begin{equation}
\Delta \chi^2_\nu = \chi^2_\nu - \chi^2_{\nu,{\rm smooth}} > 0.4
\end{equation}
where $\chi^2_\nu$ is the reduced $\chi^2$ of the original light curve
with respect to a constant model, and $\chi^2_{\nu,{\rm smooth}}$ is
the reduced $\chi^2$ of the light curve with the smoothed, phase-folded
version subtracted.  Again, $\chi^2_{\nu,{\rm smooth}}$ was calculated
for the three observing runs separately, and then summed, to account
for any evolution in amplitude and/or phase.  Since fitting for
separate $m_{\rm dc}$ coefficients in each observing run renders the
period search insensitive to very long-period modulations appearing as
simple ``DC offsets'' between the runs, we evaluated $\chi^2_\nu$ for
a three constant model, where we allowed for a different constant
light curve level in each observing run.

The threshold of $\Delta \chi^2_\nu > 0.4$ was used for the M34 data
and appears to work well here too, checked by examining all of the
light curves for two of the detectors, chosen randomly.  A total of
$1700$ objects were selected by this automated part of the procedure.
These light curves were then examined by eye, to define the final
sample of periodic variables.  A total of $812$ light curves passed
this final stage, where the remainder had spurious variability (caused
by systematic effects) or were too ambiguous to be included.

\subsection{Simulations}
\label{sim_section}

Simulations were performed following the method detailed in
\citet{i2006}, injecting simulated signals of $2$ per cent amplitude
and periods chosen following a uniform distribution on $\log_{10}$
period from $0.1$ to $20\ {\rm days}$, into light curves covering a
uniform distribution in mass, from $1.0$ to $0.1\ \msun$.  A total of
$2111$ objects were simulated.  The phase of the modulations was
randomised for each observing run in order to provide a more realistic
evaluation of the sensitivity of our method across changes in spot
patterns.

\begin{figure*}
\centering
\includegraphics[angle=270,width=6in]{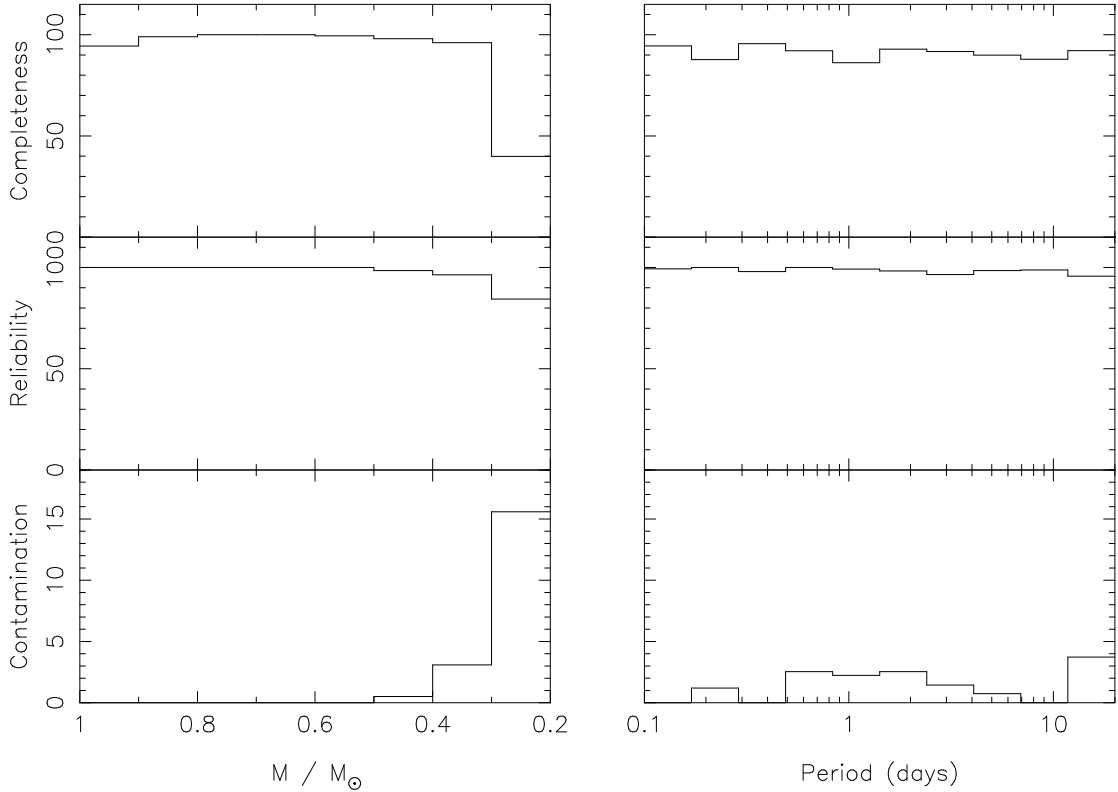}

\caption{Results of the simulations for $0.02\ {\rm mag}$ amplitude
  expressed as percentages, plotted as a function of mass (left) and
  period (right).  The simulated region covered $0.1 < {\rm M}/\msun <
  1.0$ in order to be consistent with the M50 sample.  {\bf Top
  panels}: completeness as a function of real (input) period.  {\bf
  Centre panels}: Reliability of period determination, plotted as the
  fraction of objects with a given true period, detected with the
  correct period (defined as differing by $< 20$ per cent from the
  true period).  {\bf Bottom panels}: Contamination, plotted as the
  fraction of objects with a given detected period, having a true
  period differing by $> 20$ per cent from the detected value.}

\label{sim_results}
\end{figure*}

The results of the simulations are shown in Figure \ref{sim_results}
as diagrams of completeness, reliability and contamination
as a function of period and stellar mass.  Broadly, our period
detections are close to $100$ per cent complete for these amplitudes
from $1.0\ \msun$ down to $0.3\ \msun$, with remarkably little period
dependence.  At the very lowest masses (particularly, in the $0.2 <
M/\msun < 0.3$ bin of the diagram), the completeness drops
substantially, and the reliability of the detected periods
deteriorates slightly, due to the increased noise level in the light
curves.  For $M < 0.2\ \msun$, the completeness drops essentially to
zero for $2\%$ modulations.

\begin{figure}
\centering
\includegraphics[angle=270,width=3in,bb=59 107 581 630,clip]{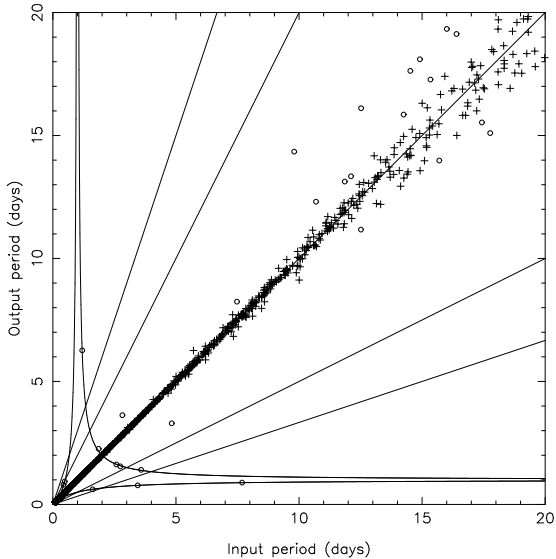}

\caption{Detected period as a function of actual (input) period for our
  simulations.  Objects plotted with crosses had fractional period
  error $< 10$ per cent, open circles $> 10$ per cent.  The straight
  lines represent equal input and output periods, and factors of $2$,
  $3$, $1/2$ and $1/3$.  The curved lines are the loci of the $\pm 1\
  {\rm day^{-1}}$ aliases resulting from gaps during the day.  The
  majority of the points fall on (or close to) the line of equal
  periods.}

\label{periodcomp}
\end{figure}

Figure \ref{periodcomp} shows a comparison of the detected
periods with real periods for our simulated objects, indicating good
reliability of period recovery, with only a small amount of aliasing,
especially when compared to some of our previous data-sets (e.g. M34;
\citealt{i2006}), presumably due to the extended time-coverage of the
M50 observations.

\section{Results}
\label{results_section}

\begin{figure}
\centering
\includegraphics[angle=270,width=3.5in]{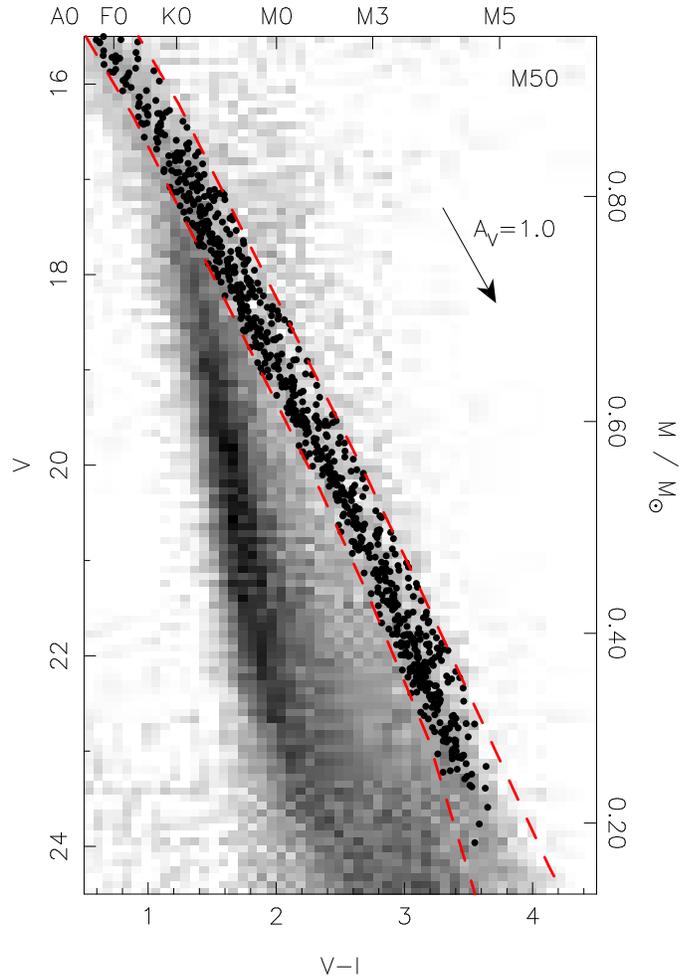}

\caption{Magnified $V$ versus $V - I$ CMD of M50, for objects
with stellar morphological classification, as Figure \ref{cmd},
showing all $812$ candidate cluster members with detected periods
(black points).  The dashed lines show the cuts used to select
candidate cluster members (see \S \ref{cmd_section}).}

\label{cands_on_cmd}
\end{figure}

The locations of our detected periodic variable candidate cluster
members on a $V$ versus $V-I$ CMD of M50 are shown in Figure
\ref{cands_on_cmd}.  The diagram indicates that the majority of
the detections lie on the single-star cluster sequence, as would
be expected for rotation in cluster stars as opposed to, say,
eclipsing binaries.

The properties of all our rotation candidates are listed in Table
\ref{cand_table}.

\begin{table*}
\centering
\begin{tabular}{lllrrrrrrrrrrrrrrr}
\hline
Identifier    &RA    &Dec   &$V$   &$\sigma(V)$ &$I$ &$\sigma(I)$ &$P$ &$\alpha_1$ &$\sigma(\alpha_1)$ &$\alpha_2$ &$\sigma(\alpha_2)$ &$\alpha_3$ &$\sigma(\alpha_3)$ &$M$ &$R$ \\
              &J2000 &J2000 &(mag) &(mag) &(mag) &(mag) &(days) &(mag)      &(mag)      &(mag)      &(mag)      &(mag)      &(mag)      &($\msun$) &($\rsun$) \\
\hline
M50-1-98    &07 01 35.53 &-08 33 10.6 &17.082 &0.001 &15.911 &0.001 &12.807 &0.0091 &0.0003 &0.0035 &0.0002 &0.0115 &0.0011 &0.81 &0.76 \\
M50-1-413   &07 01 38.56 &-08 34 35.7 &21.993 &0.015 &18.950 &0.007 & 0.741 &0.0078 &0.0010 &0.0209 &0.0011 &0.0229 &0.0012 &0.37 &0.36 \\
M50-1-651   &07 01 40.52 &-08 33 28.3 &16.521 &0.001 &15.423 &0.001 & 6.811 &0.0055 &0.0002 &0.0063 &0.0002 &0.0024 &0.0004 &0.88 &0.84 \\
M50-1-979   &07 01 43.77 &-08 32 58.5 &21.901 &0.014 &18.710 &0.006 & 0.865 &0.0189 &0.0009 &0.0108 &0.0009 &0.0104 &0.0014 &0.41 &0.39 \\
M50-1-1130  &07 01 45.15 &-08 34 11.0 &21.330 &0.009 &18.538 &0.006 & 0.924 &0.0138 &0.0014 &0.0173 &0.0008 &0.0292 &0.0013 &0.44 &0.41 \\
\hline
\end{tabular}

\caption{Properties of our $812$ rotation candidates, including $V$
  and $I$-band magnitudes and uncertainties from our CCD photometry,
  the period $P$, $i$-band amplitudes $\alpha_1$, $\alpha_2$,
  $\alpha_3$ and uncertainties (magnitudes, in the instrumental
  bandpass), interpolated mass and radius (from the models of 
  \citealt{bcah98}, derived using the $I$ magnitudes).  In the table,
  $\sigma(x)$ denotes the uncertainty in quantity $x$.  Note that
  these uncertainties do not incorporate systematic errors (e.g. in
  the zero point calibration), that dominate at the bright end.  Our
  identifiers are formed using a simple scheme of the cluster name,
  CCD number and a running count of stars in each CCD, concatenated
  with dashes.  The full table is available in the electronic edition.
  Machine readable copies of the data tables from all the Monitor
  rotation period publications are also available at {\tt
  http://www.ast.cam.ac.uk/research/monitor/rotation/}.}
\label{cand_table}
\end{table*}

\subsection{M50 rotation periods}
\label{prv_section}

\begin{figure}
\centering
\includegraphics[angle=270,width=3in]{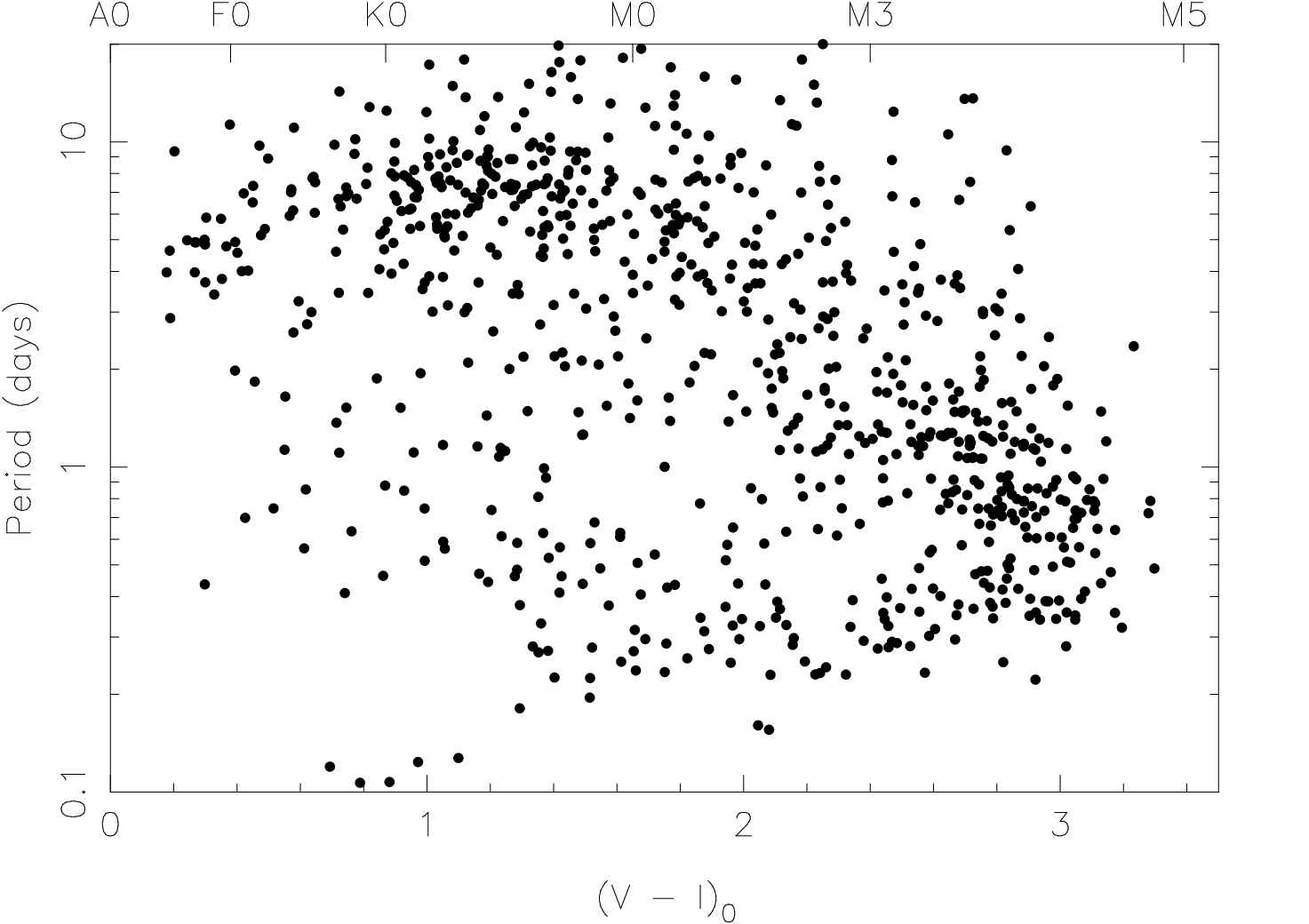}
\includegraphics[angle=270,width=3in]{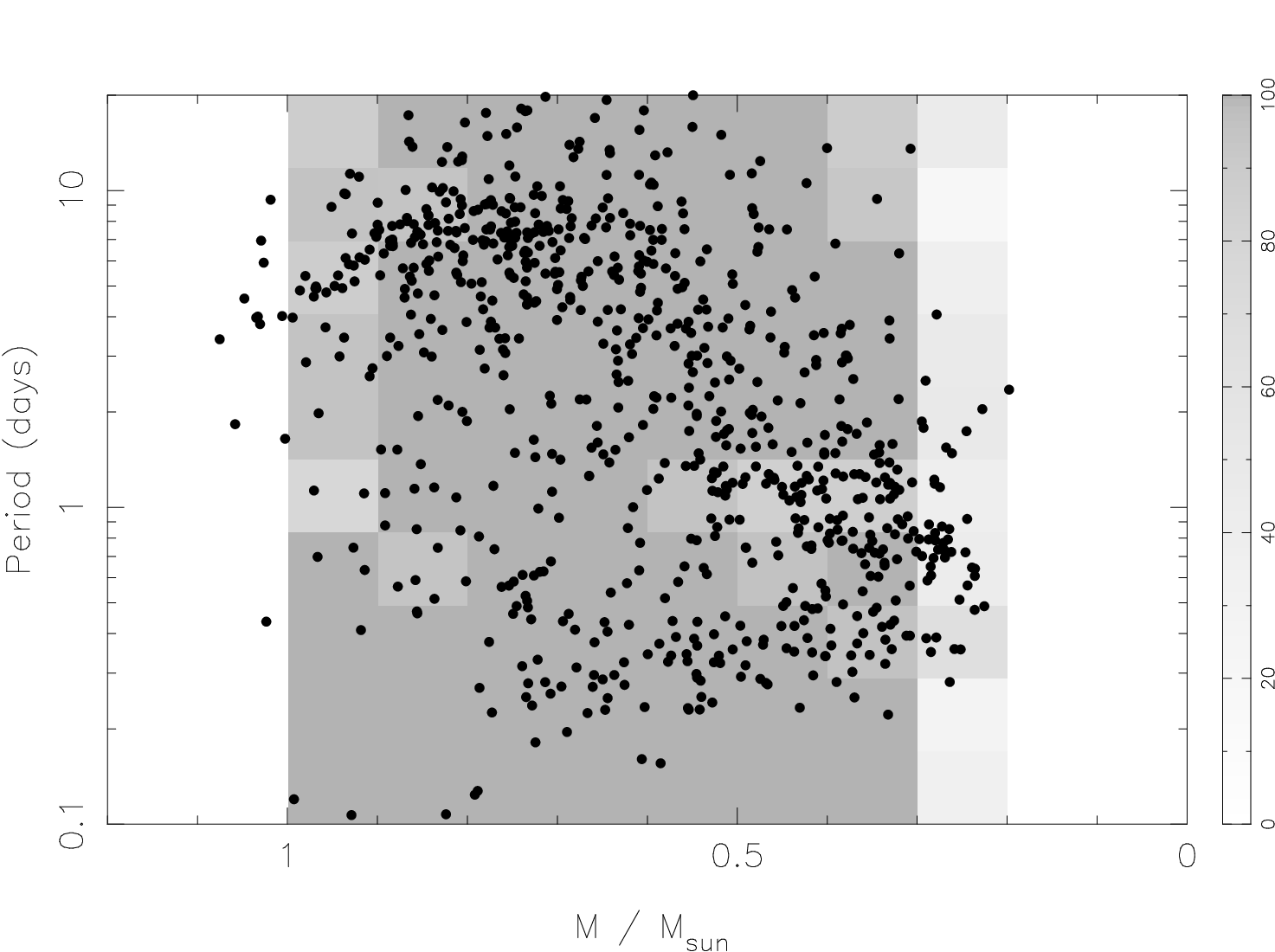}

\caption{Plots of rotation period as a function of dereddened $V-I$
  colour (top), and mass (bottom) for M50, deriving the masses
  using the $130\ {\rm Myr}$ NextGen mass-magnitude relations of
  \citet{bcah98} and our measured $I$-band magnitudes.  In the lower
  diagram, the greyscales show the completeness for $0.02\ {\rm mag}$
  periodic variations from the simulations.}

\label{pcd}
\end{figure}

Plots of period as a function of $V-I$ colour and mass for the 
photometrically-selected candidate cluster members are shown in
Figure \ref{pcd}.  Below $\sim 0.7\ \msun$ (or M0), these diagrams
reveal a correlation between stellar mass (or spectral type) and the
longest rotation period seen at that mass, with a clear lack of slow
rotators at very low masses.  This trend is also followed by the
majority of the rotators in this mass range, with only a tail of
faster rotators to $\sim 0.2\ {\rm day}$ periods, and very few objects
rotating faster than this.  This is very similar to what we found in
the earlier NGC 2516 and NGC 2547 studies (\citealt{i2007b};
\citealt{i2008a}).

Above $\sim 0.7\ \msun$, the M50 data indicate an inverse trend, of
decreasing rotation period (faster rotation) as a function of
increasing mass.  The slope of this relation is much shallower than
the one below $0.7\ \msun$.  The existence of such a relation was
noted by \citet{hartman2008} in M37 ($\sim 550\ {\rm Myr}$), and is
also clearly evident in the $v \sin i$ data for NGC 2516 and the
Pleiades, and recently published rotation period data in M35
\citep{meibom08}, as shown in Figure \ref{comp_m50m35n2516} (left
panel), which will be discussed in \S \ref{comp_section}.

These morphological features do not appear to be a result of sample
biases.  In particular, the simulations show that the survey is
sensitive to much shorter periods than $0.2\ {\rm day}$, and the upper
limit in detectable periods is not mass-dependent, so this cannot
explain the morphology of the upper envelope of rotation periods in
Figure \ref{pcd}.  Moreover, the presence of these features in
multiple rotation period data-sets with differing selection biases,
and in $v \sin i$ data, strongly indicates that they are real features
of the underlying rotation rate distribution of low-mass stars.

\begin{figure}
\centering
\includegraphics[angle=270,width=3in]{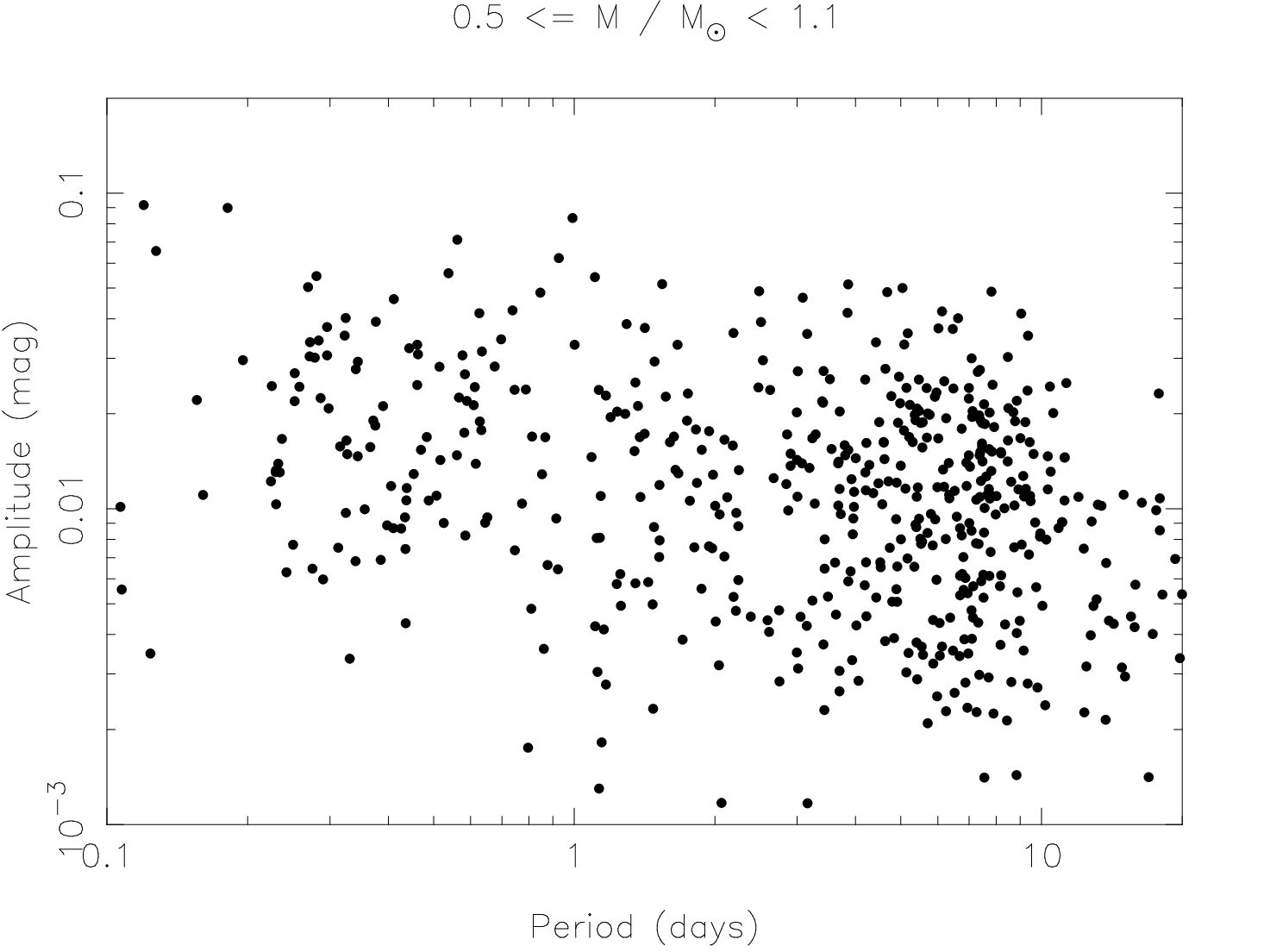}
\includegraphics[angle=270,width=3in]{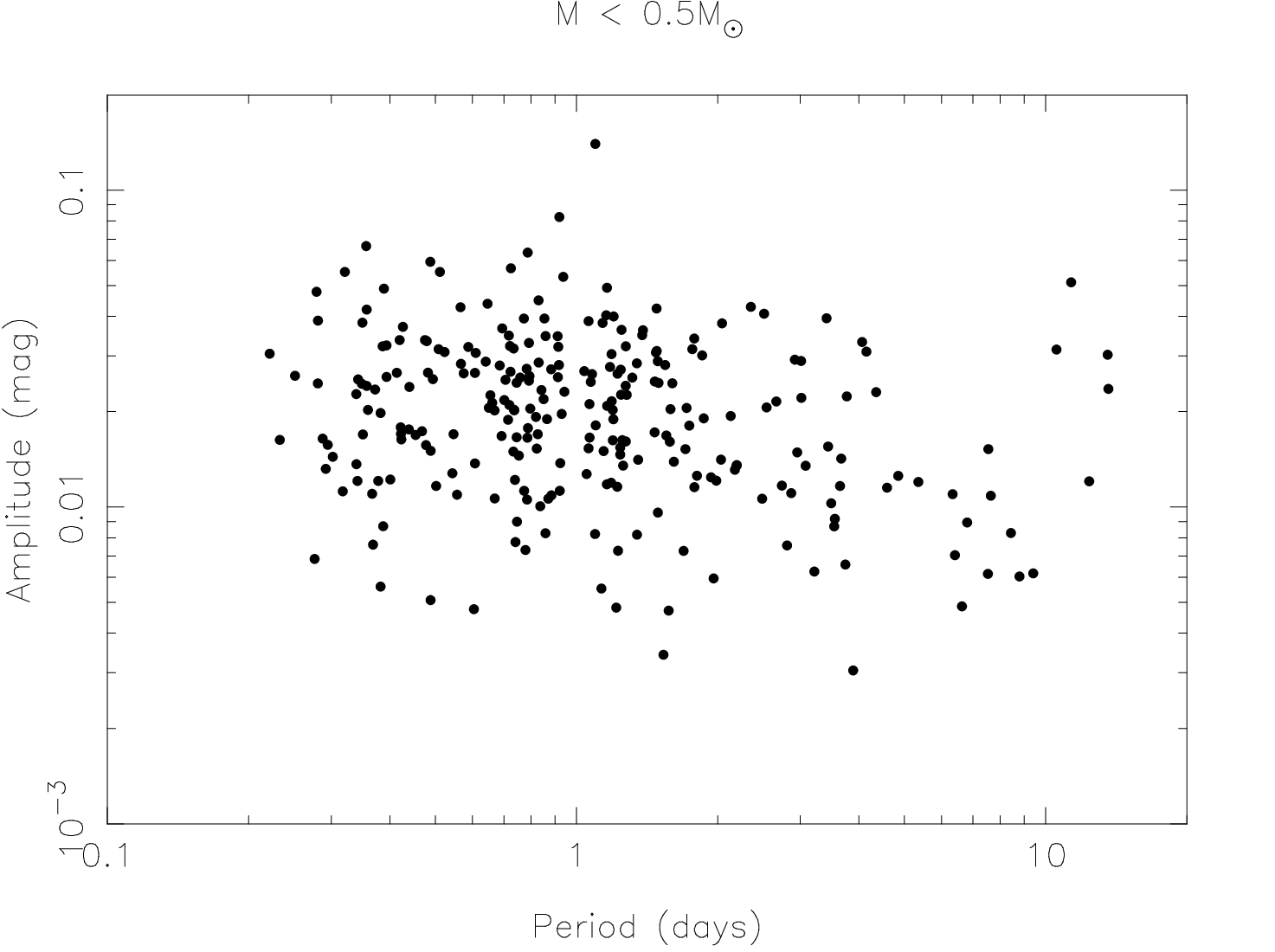}

\caption{Plot of amplitude as a function of period for M50 in two
  mass bins: $0.5 \le M/\msun < 1.1$ (top) and $M < 0.5\ \msun$
  (bottom).}

\label{pad}
\end{figure}

As a further check, we can examine the distribution of rotation
periods as a function of amplitude.  This is shown in Figure
\ref{pad}, split into two mass bins.  There does not appear to be any
clear sensitivity bias, with very low amplitudes being detected over
the full range of period, for the $M < 0.5\ \msun$ bin.

The higher-mass bin ($0.5 \le M/\msun < 1.1$) appears to show a
correlation between amplitude and rotation period, in the sense that
longer-period objects show smaller amplitudes.  This is confirmed by
applying a non-parametric Spearman rank correlation analysis, which
gave $r_s = -0.25$, corresponding to a probability of $3 \times
10^{-9}$ that the quantities are uncorrelated given the sample size.
This trend is also present in the lower-mass bin, but at a lower
significance level (mostly due to the smaller sample size), with $r_s
= -0.22$, and a probability of $4 \times 10^{-4}$.

This correlation could be the result of a bias against detection of
very small-amplitude modulations at short periods, but this is in the
opposite sense to the expected trend: it is easier to detect small
modulations at short periods since the slope of the light curve within
each night is larger, which contributes significantly to our
confidence that the variation is real.  We speculate that there are
two likely causes of the trend we have observed: either this is a
result of the well-known relation between rotation rates and stellar
activity (e.g. see \citealt{g2004} for a review), in the sense that
fast rotators are more active, and therefore are more spotted, giving
rise to a larger photometric amplitude; or, the contamination of our
sample by field stars, which are in general less active and more
slowly rotating than cluster stars, leads to a population of
contaminant objects at the long-period end giving rise to the apparent
correlation.  The upper panel of Figure \ref{pad} already contains a
hint that the latter may be the case: the distribution of amplitudes
for periods $> 10\ {\rm days}$ appears to be systematically skewed
toward smaller amplitudes than that for periods $< 10\ {\rm days}$.
Moreover, we expect there to be very few cluster stars rotating more
slowly than $10\ {\rm days}$ at this age from rotational evolution
models (e.g. see \citealt{i2007b}; \citealt{i2008a};
\citealt{i2008b}).

\subsection{Radial distribution of periodic variables}

\begin{figure}
\centering
\includegraphics[angle=270,width=3in]{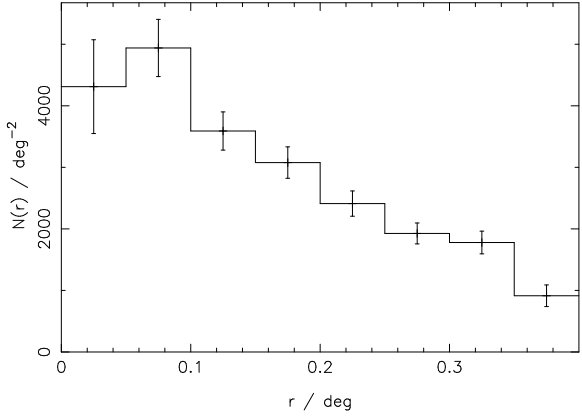}

\caption{Radial distribution for objects with detected periods.  The
  counts in each annulus have been corrected for the area of sky
  covered by the CCD mosaic by summing the area of all the pixels in
  falling within that annulus.  This correction is only approximate
  for the outermost regions (e.g. $r \ga 0.35^\circ$) since it does
  not take into account incompleteness in the detections close
  to the edges of the detector, so the counts in the $0.35^\circ < r <
  0.4^\circ$ bin in particular may be unreliable.  The apparent
  deficit of objects in the innermost ($r < 0.05^\circ$) bin is most
  likely due to a small error in the position of the cluster centre.}

\label{radial_dist_periodic}
\end{figure}

\begin{figure}
\centering
\includegraphics[angle=270,width=3in]{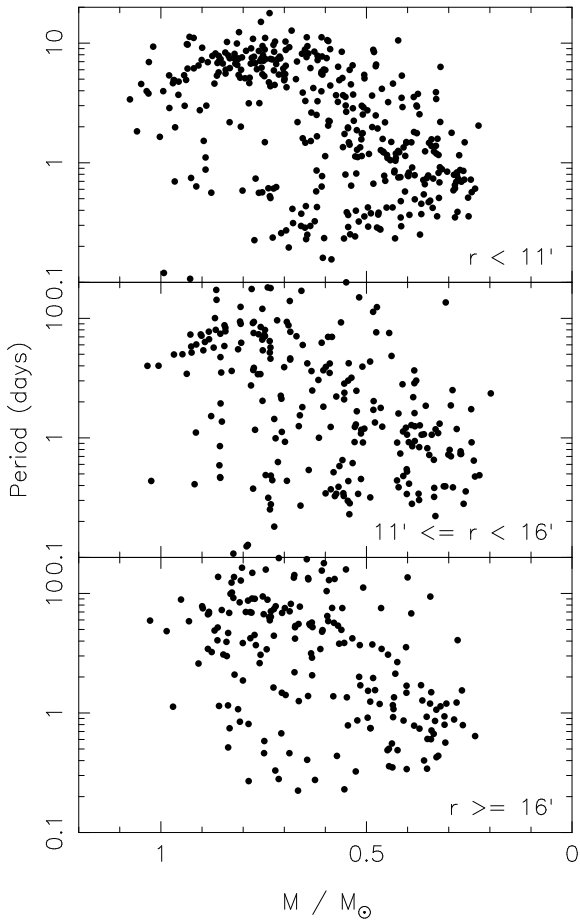}

\caption{Rotation period as a function of mass, as Figure \ref{pad},
  plotted in three radial bins of approximately equal area on the sky:
  $r < 11'$ (top), $11' \le r < 16'$ (centre), and $r \ge 16'$
  (bottom).  The latter bin extends to the edge of the $\sim 36'
  \times 36'$ field of view of our observations.}

\label{radial_pmd}
\end{figure}

In order to further examine the effect of field contamination, we show
in Figure \ref{radial_dist_periodic} the radial distribution of the
detected periodic variable objects, and in Figure \ref{radial_pmd} the
rotation period distribution as a function of mass in three radial
bins, measuring the radius $r$ with respect to the approximate
position of the cluster centre.  We adopt a position of R.A. $7^{\rm
h}02^{\rm m}47.4^{\rm s}$, Dec. $-8^\circ20'43''$ (\citealt{sharma06};
see their table 5).  These authors derive a cluster core radius of
$6.5'$ and a total extent of $17'$, which are broadly consistent with
Figure \ref{radial_dist_periodic}, so we expect field stars to
dominate our largest-radius bin in Figure \ref{radial_pmd}.  While
this is clearly not the case, given the presence of the same
well-defined morphology in rotation period as a function of mass in
all three bins (presumably due to the cluster population), the
fraction of objects with periods $> 10\ {\rm days}$ does appear to
increase with $r$, becoming especially clear in the bottom panel of
the figure.  This suggests that these objects may indeed be
slowly-rotating field stars contaminating the sample.

The existence of a significant population of cluster stars in the
lower panel of Figure \ref{radial_pmd} is not necessarily inconsistent
with the conclusions of \citet{sharma06}.  Firstly, by requiring the
objects to have measured rotation periods, we strongly bias the sample
in favour of cluster stars, since these are more active.  Secondly,
many clusters are found to have a larger apparent extent at low masses
(e.g. \citealt{larson82}; \citealt{mcnamara86}; \citealt{sagar88}), in
some cases even having a ``halo'' population of low-mass stars (e.g. h
and $\chi$ Persei: \citealt{schild67}; \citealt{currie07}).  Our
rotation period sample probes $\sim 3\ {\rm mag}$ deeper in $V$-band
than the survey by \citet{sharma06}, and is therefore dominated by
lower-mass stars.  The radial distribution shown in the lowest-mass
bin of their Figure 4 supports this hypothesis, showing a larger
cluster population at large radii than their higher-mass bins.  Due to
the limited angular coverage of the present survey, it is difficult to
confirm this from Figure \ref{radial_dist_periodic}.

\subsection{Comparison with NGC 2516 and M35}
\label{comp_section}

\begin{figure*}
\centering
\includegraphics[angle=270,width=7in]{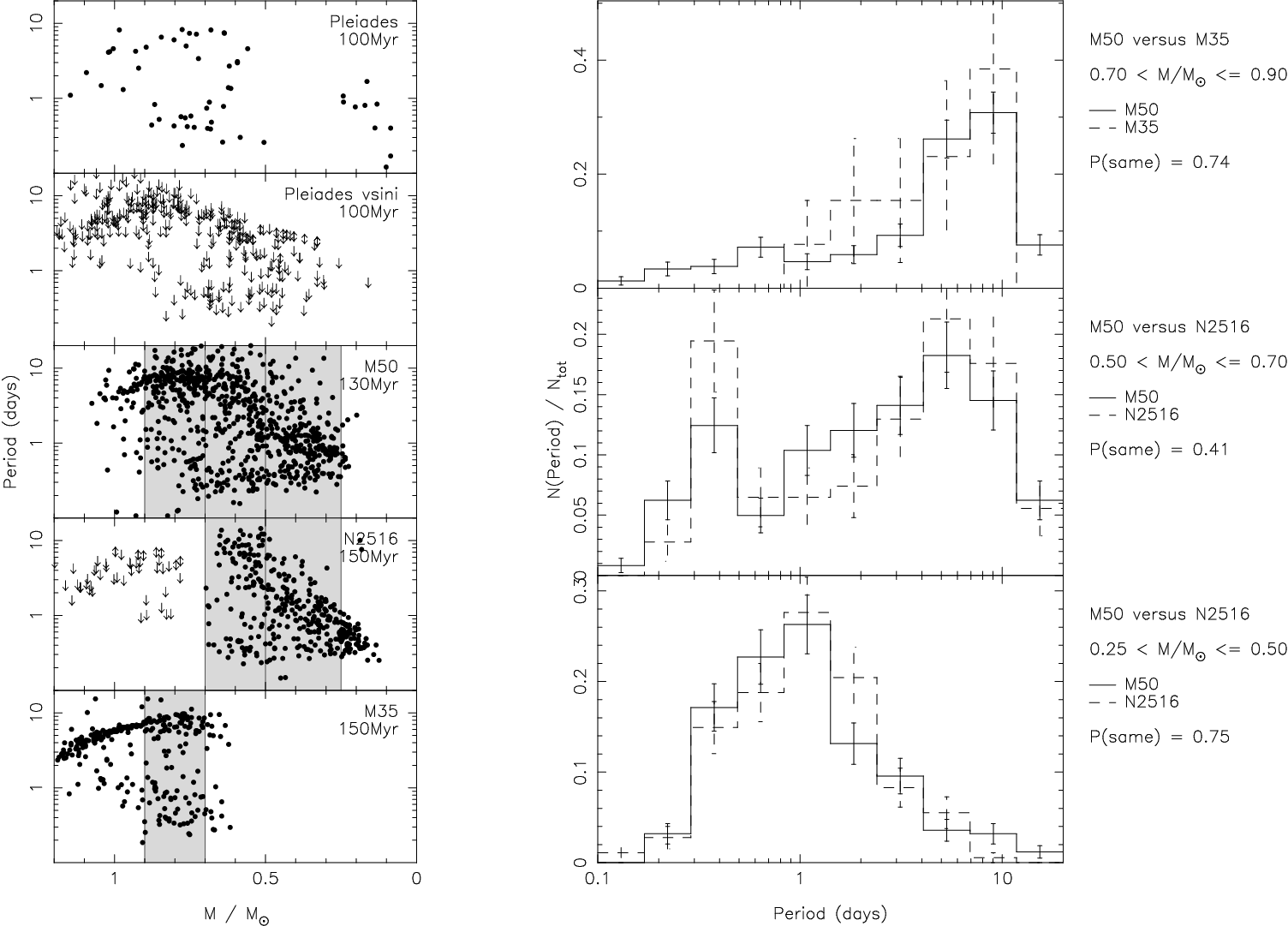}

\caption{Comparison of the M50 rotation period distribution with
  rotation period measurements in the Pleiades (\citealt{ple1};
  \citealt{ple2}; \citealt{ple3}; \citealt{ple4}; \citealt{ple5};
  \citealt{ple6}; \citealt{ple7}; \citealt{t99}; \citealt{se2004b}),
  $v \sin i$ measurements in the Pleiades (\citealt{plev1};
  \citealt{plev2}; \citealt{plev3}; \citealt{plev4}; \citealt{plev5};
  \citealt{plev6}), and rotation periods in NGC 2516 and M35.  {\bf
  Left:} rotation period as a function of mass for all five samples.
  Masses were computed using the measured $I$-band magnitudes,
  interpolating the models of \citet{bcah98} to the appropriate
  metallicity and age for each cluster, using the estimates of these
  parameters from the literature quoted in sections
  \ref{intro_section} and \ref{comp_section} (for M50, we simply
  assumed solar metallicity given the lack of an estimate).
  {\bf Right:} comparison of histograms of rotation period binned in
  mass to those regions where the cluster samples have sufficient
  overlap for a meaningful comparison.  The grey shaded regions in the
  left panels indicate the mass ranges used for the histograms on the
  right.  Note that these distributions have not been corrected for
  contamination by field stars due to the lack of follow-up data.
  Nevertheless, the rotation period distributions in all these clusters
  are qualitatively very similar, and appear to be statistically
  indistinguishable given the present sample sizes.}

\label{comp_m50m35n2516}
\end{figure*}

By comparing the rotation period distributions in different clusters
of the same age, we can begin to constrain the effect of cluster
environment, and other parameters such as metallicity, on the
distribution of rotation periods.  In the case of M50, suitable
rotation period samples are available in the literature for NGC 2516
\citep{i2007b} and M35 \citep{meibom08}, with a reasonable overlap in
the mass range covered.  The metallicity of M50 does not yet appear to
have been reported in the literature, but estimates are available for
NGC 2516 indicating that it is near-solar (${\rm [Fe/H]} = -0.05 \pm
0.14$; \citealt{ter2002}), and for M35 indicating a sub-solar
metallicity (${\rm [Fe/H]} = -0.21 \pm 0.10$; \citealt{barrado01}).
Figure \ref{comp_m50m35n2516} shows a comparison of the rotation
periods measured in these two clusters with the M50 distribution from
the present study.

Examining the left-hand panels of Figure \ref{comp_m50m35n2516} first,
the morphology displayed in all the clusters is clearly very similar
over the mass ranges in common.  The Pleiades $v \sin i$ data appear
to show a slightly different ``turn-over'' point for the slowest
rotators, at $\sim 0.9\ \msun$ rather than $0.7\ \msun$ for the other
clusters, but this may be the result of the $\sin i$ ambiguity and the
difficulty of detecting the slowest rotating objects via this method.
Therefore, qualitatively, the distributions of rotation period versus
mass for all the clusters appear to be very similar.

In order to allow a quantitative comparison to be made, we have
compared the distributions of rotation periods in three mass bins,
chosen to resolve morphological changes in the diagram while
concentrating on the regions where the various samples overlap
sufficiently to allow a meaningful comparison.  These are shown in the
right-hand panels of Figure \ref{comp_m50m35n2516}.  In all three
cases, the distributions are extremely similar, generally differing by
less than their combined respective Poisson counting uncertainties.
This was confirmed by applying non-parametric two-sided
Kolmogorov-Smirnov tests to the distributions, the results of which
are given in the figure.  In all cases the distributions are
statistically indistinguishable.

The present samples in these three clusters therefore do not show
evidence for dependence of rotation on cluster environment, or a
``third parameter'' such as metallicity.  This is of course not
conclusive evidence that such a dependence does not exist, especially
since the environment and other parameters are relatively similar for
the three clusters considered in this work.  In order to conduct a
more stringent test, additional rotation period observations in a
substantially different environment at similar age will be needed.

\subsection{Changes in amplitude: evolution of the spot patterns}

The availability of three epochs of data corresponding to the three
different observing runs allows us to constrain the evolution of the
spot patterns giving rise to the photometric modulations, by examining
the evolution of the amplitude between the runs.  Moreover, by
comparing results in multiple clusters (e.g. our NGC 2362 data-set,
which has the same sampling), we can examine the dependence of the
spot evolution rate on cluster age.

\begin{figure}
\centering
\includegraphics[angle=270,width=3in]{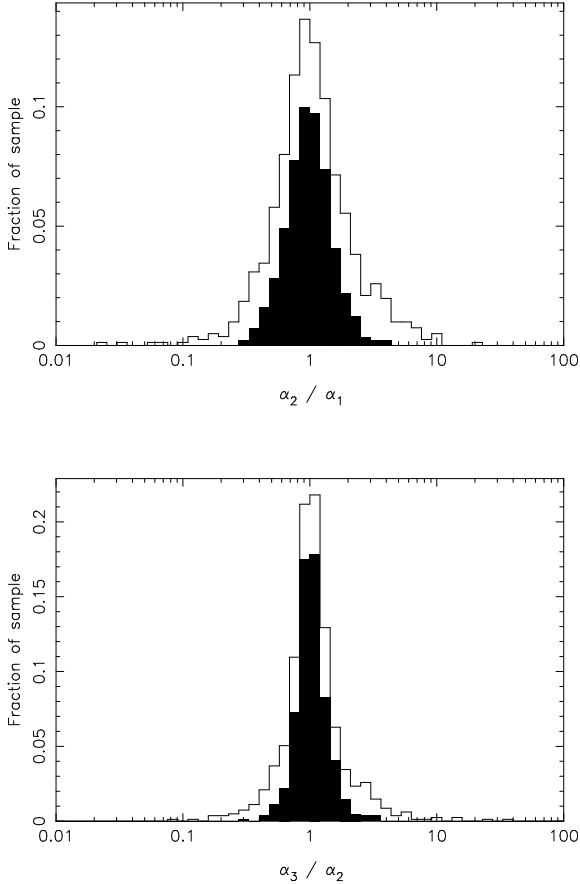}

\caption{Histograms of the relative amplitudes from the sine curve
  fitting procedure described in \S \ref{method_section} for the
  second and first observing runs (upper panel), and the third and
  second observing runs (lower panel).  The mid-points of these were
  separated by $\sim 10\ {\rm months}$ and $1\ {\rm month}$
  respectively.  The open histogram shows the entire sample, and the
  solid histogram only those objects with amplitudes $> 1\%$.  The
  standard deviations of the distributions, in $\log_{10}$ units, for
  the open (solid) histograms are $0.25$ ($0.17$) for the upper panel,
  and $0.14$ ($0.10$) for the lower panel.}

\label{evol_amp}
\end{figure}

Figure \ref{evol_amp} shows histograms of the ratio of the amplitudes
between pairs of observing runs.  As expected, over the interval
between the first two runs ($\sim 10\ {\rm months}$), compared to that
between the second two runs ($\sim 1\ {\rm month}$), the standard
deviation of the amplitude ratio is larger, by a factor of $\sim 1.8$,
indicating that a larger fraction of the amplitudes evolved over the
longer time-span between the first two observing runs.  This clearly
demonstrates that the spot patterns on our target stars must have
evolved over the $10\ {\rm month}$ gap.  The evolution in amplitude
is, nonetheless, relatively modest, which is not surprising since it
is reasonable to expect stars of a given activity level to maintain
comparable levels of spot coverage over long time-scales, with the
phase of the modulations evolving according to the short time-scale
evolution of individual spots or spot groups.

\begin{figure}
\centering
\includegraphics[angle=270,width=3in]{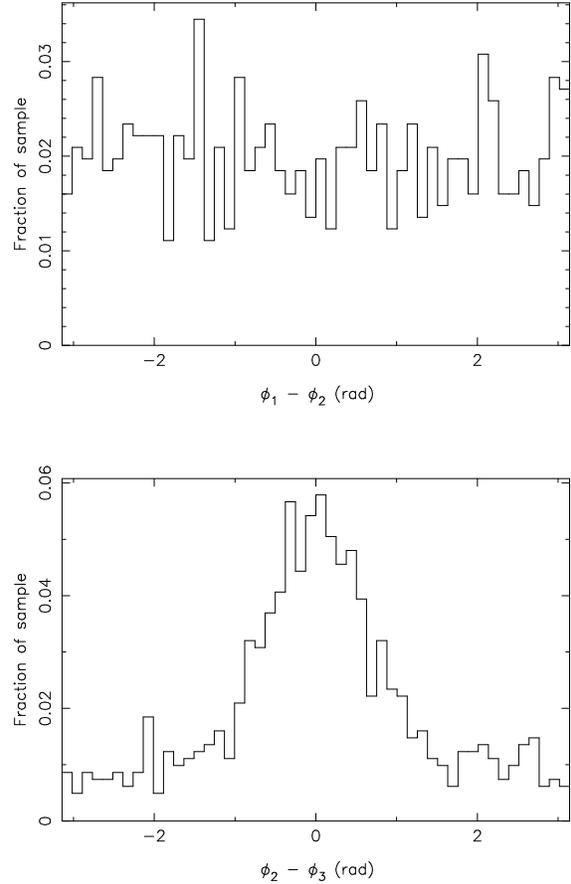}

\caption{Histograms of the relative phase from the sine curve
  fitting procedure described in \S \ref{method_section} for the
  second and first observing runs (upper panel), and the third and
  second observing runs (lower panel).  The upper distribution is
  consistent with being uniform, indicating that the phase is essentially
  randomised over the $10\ {\rm month}$ time-span between the first and
  second observing runs.  The standard deviation of the lower
  distribution is $\sim 1.0\ {\rm rad}$.}

\label{evol_phi}
\end{figure}

Unfortunately, for the sampling strategy we have used, we cannot use
the phase information to examine this evolution.  This arises because
the number of cycles of modulation over the $\sim 10\ {\rm month}$ gap 
between the first and second observing runs is large, and unknown.
Therefore, small errors in the rotation period, accumulated over the
many cycles occurring in the $10\ {\rm month}$ gap, can lead to large
phase shifts between the two observing runs.  For example, even a one
per cent error in a rotation period of $10\ {\rm days}$ gives rise to
a cumulative uncertainty of $\sim 0.3\ {\rm cycle}$ after $10\ {\rm
  months}$, and this effect is worse for shorter periods.  Figure
\ref{evol_phi} illustrates this effect, showing that over the gap
between the first and second observing runs, the phase is essentially
randomised for $100\%$ of the sample.

Therefore, with the present data-set, it is difficult to further
constrain the spot lifetimes giving rise to the observed evolution.

\section{Conclusions}
\label{conclusions_section}

We have reported on results of an $i$-band photometric survey of M50,
covering $\sim 0.4\ {\rm sq. deg}$ of the cluster.  Selection
of candidate members in a $V$ versus $V-I$ colour-magnitude diagram
using an empirical fit to the cluster sequence found $4249$
candidate members, over a $V$ magnitude range of $15.5 < V < 26$
(covering masses in the range $0.1 \la M/\msun \la 1.1$).  The likely
field contamination level was estimated using a simulated catalogue of
field objects from the Besan\c{c}on Galactic models \citep{r2003},
finding an overall contamination level of $\sim 47$ per cent, implying
that there are $\sim 2300$ real cluster members over this mass range
in our field-of-view.

We derived light curves for $\sim 63\,000$ objects in the M50
field, achieving a precision of $< 1$ per cent per data point over $15
\la i \la 19$.  The light curves of our candidate cluster members were
searched for periodic modulations, presumably due to stellar rotation,
giving $812$ detections over the mass range $0.2 < M/\msun < 1.1$.

The rotation period distribution as a function of mass was found to
show a clear mass-dependent morphology, statistically
indistinguishable from the distributions in the literature for
M35 and NGC 2516 once the different mass ranges probed by the surveys
are taken into account.  Thus the M50 results do not yet provide any
indication of a dependence of rotation rates on cluster environment,
or a ``third parameter'' such as metallicity.

Finally, we demonstrated evidence that the photometric amplitudes of
a significant fraction of our targets do indeed evolve over timescales
of $\sim 10\ {\rm months}$, as expected (for example, by analogy with
the Sun).

\section*{Acknowledgments}

Based on observations obtained at Cerro Tololo Inter-American
Observatory, a division of the National Optical Astronomy
Observatories, which is operated by the Association of Universities
for Research in Astronomy, Inc. under cooperative agreement with the
National Science Foundation.  This research has made use of the SIMBAD
database, operated at CDS, Strasbourg, France, and the WEBDA database,
operated at the Institute for Astronomy of the University of Vienna.
The Open Cluster Database, as provided by C.F. Prosser and
J.R. Stauffer, may currently be accessed at {\tt
  http://www.noao.edu/noao/staff/cprosser/}, or by anonymous ftp to
{\tt 140.252.1.11}, {\tt cd /pub/prosser/clusters/}.

JI gratefully acknowledges the support of a PPARC studentship, and SA
the support of a PPARC postdoctoral fellowship, during the time the
majority of the work was carried out.  We thank the referee for his
comments, which have helped to improve the paper.

\end{document}